\documentclass[prl,aps,twocolumn,floatfix]{revtex4}
\usepackage{graphicx, amsmath}
\begin{document}

\title{Nature of excitations of the 5/2 fractional quantum Hall effect}
\author{Csaba T\H oke$^{*}$}
\author{Nicolas\ Regnault$^{\dagger}$}
\author{Jainendra K.\ Jain$^{*}$}
\affiliation{$^{*}$Department of Physics, 104 Davey Laboratory, The Pennsylvania State University, Pennsylvania, 16802}
\affiliation{$^{\dagger}$ Laboratoire Pierre Aigrain, D\'epartement de Physique, 24 rue Lhomond, 75005 Paris, France}

\begin{abstract}
It is shown, with the help of exact diagonalization studies on systems with up to sixteen electrons, in the presence of up to two delta function impurities, that the Pfaffian model 
is inadequate for the actual quasiholes and quasiparticles of the 5/2 fractional 
quantum Hall effect.   Implications for non-Abelian statistics are discussed.
\end{abstract}
\maketitle

Comparisons with exact diagonalization studies on finite size systems have 
served as a litmus test for theoretical proposals on fractional quantum Hall effect.  
This paper reports on such tests for the Pfaffian model of the 5/2 FQHE, 
which describes a paired state of composite fermions \cite{Moore1,GWW,Greiter}.  
The Pfaffian wave function for the 
incompressible ground state has been studied before; the focus in this paper is 
on quasiparticles and quasiholes of this state.   
A case has been made, both from analytical arguments
\cite{Moore1,Nayak1,Read1} and numerical calculations \cite{TSimon}, that 
the Pfaffian quasiholes exhibit non-Abelian 
braiding statistics.   However, the relevance of this result to the actual Coulomb 
quasiholes has not been established, and there has been a debate in the literature on 
whether the Coulomb state is adiabatically connected to the Pfaffian 
state \cite{Ho95}.  This question is interesting in its own right, and also 
because of recent proposals of exploiting non-trivial braiding properties of certain 
FQHE quasiparticles for quantum computation. We investigate below how well the
Pfaffian wave functions represent the solutions of the Coulomb interaction, with and 
without the presence of weak impurities.

Our calculations are performed in the standard spherical geometry
\cite{Haldane,Wu}, 
with $N$ electrons on the surface of a sphere subjected to a magnetic field 
that produces a flux $2Q\phi_0$ through the surface.  
Here  $\phi_0=hc/e$ and $2Q$ is an integer.
Only electrons in the second Landau level (LL) are considered, assumed to be 
fully spin polarized.  The eigenstates are labeled by an orbital angular momentum.

The system of Coulomb electrons at $\nu=1/2$ in the second LL 
is equivalent to the system of lowest-LL electrons at $\nu=1/2$ 
interacting via an effective interaction.
The Pfaffian wave function in the lowest LL is given by 
(Moore and Read \cite{Moore1})
\begin{equation}
\label{pfaffsphere}
\Psi^{\rm Pf}_0=\text{Pf}\left(\frac{1}{u_iv_j-v_iu_j}\right)\prod_{i<j}(u_iv_j-v_iu_j)^2\;,
\end{equation}
where $u_i=\cos\frac{\theta_i}{2}e^{-i\phi_i/2}$ and 
$v_i=\sin\frac{\theta_i}{2}e^{i\phi_i/2}$.  
The wave function for two quasiholes
at $(U_1,V_1)$ and $(U_2,V_2)$ is given by \cite{Moore1}
\begin{equation}
\Psi^{\rm Pf}_{\rm 2-qh}={\rm Pf}\left(M_{ij}\right)\;
\prod_{i<j}(u_iv_j-v_iu_j)^2\;,
\label{eq:pfaff12hole2n}
\end{equation} 
\begin{equation}
M_{ij}=\frac{(u_iV_1-v_iU_1)(U_2v_j-V_2u_j)+(i\leftrightarrow j)}{(u_iv_j-v_iu_j)}\;.
\end{equation} 
For two coincident quasiholes, 
$(U_1,V_1)=(U_2,V_2)\equiv(U,V)$, it reduces to a charge $1/2$ vortex, 
\begin{equation}
\Psi_V=\prod_i(u_iV-v_iU)\Psi^{\rm Pf}_0\;. 
\end{equation}
Separately, each quasihole has a charge deficiency of $1/4$ 
associated with it.  Unlike for the vortex, the density does not vanish at the position 
of a quasihole.  Analogous wave function can be written for many quasiholes. 
No simple wave functions presently exist for quasiparticles.

The above wave functions are the exact, zero-energy ground state of an 
unphysical short-range three-body interaction in the lowest LL \cite{GWW,Read1} 
\begin{equation}
H^{(3)}=(e^2/\epsilon l_B) \sum_{i<j<k} P_{ijk}(L_{\rm max})
\label{modelint}
\end{equation}
where $P_{ijk}(L_{\rm max})$ is the projection operator onto a triplet of
orbital angular momentum $L_{\rm max}=3Q-3$, 
$l_B=\sqrt{\hbar c/eB}$ is the magnetic length and 
$\epsilon$ is the static dielectric constant of the background semiconductor. 
There is no interaction when {\em two} electrons approach one another, but 
an energy cost is associated with electron triplets in their closest configuration.
When two quasiholes are present, a diagonalization of $H^{(3)}$ produces 
many zero energy states, which can be chosen to be eigenstates of $L$; 
we refer to the subspace of zero-energy wave functions as 
the ``Pfaffian quasihole (PfQH) sector."  Spatial localization of quasiholes, as 
described by the wave function $\Psi^{\rm Pf}_{\rm 2-qh}$, breaks rotational invariance, 
but the wave function still lives entirely in the PfQH sector.   The origin of non-Abelian statistics lies in the degeneracy of states in the PfQH sector, which produces, in 
general, several degenerate wave functions for a given quasihole configuration, 
thereby allowing for the possibility that 
quasihole braidings can produce different linear combinations of PfQH states, 
hence non-Abelian statistics.

\begin{table}[htb]
\begin{center}
\begin{tabular}{c|c|c|c|c|c|c|c|c}
\hline\hline
$N$ &  $L=0$ & 1 & 2 & 3 & 4 & 5 & 6 & 7 \\ 
\hline
8 & 0.64 & - & 0.48 & - & 0.52 & -  & -  & - \\
10 & - & 0.05 & - & 0.56 & - & 0.61 & -  & - \\
12 & 0.59 & - & 0.30 & - & 0.49 & - & 0.39  & - \\
14 & - & 0.39 & - & 0.13 & - & 0.39 & -  & 0.27 \\
\hline\hline
\end{tabular}
\end{center}
\caption{\label{PfQHproj2}Overlaps between the PfQH basis for two quasiholes and the lowest 
energy states for Coulomb interaction in the second Landau level at $2Q=2N-2$.
The overlaps are defined as
${\cal O}=|\langle \Psi^{(3)}_{\rm 2-qh}|\Psi^{\rm C}_{\rm 2-qh}\rangle|^2$.
The wave function $|\Psi^{\rm (3)}_{\rm 2-qh}\rangle$ at orbital angular 
momentum $L$ refers to the two quasihole eigenstate of $H^{(3)}$ 
with quantum number $L_z=L$, 
and $\left|\Psi^{\rm C}_{\rm 2-qh}\right\rangle$ is the lowest energy state  for the Coulomb interaction with the same quantum numbers.  We note that for two quasiholes, there is a single zero energy multiplet at alternate values of $L$ for $H^{(3)}$ ($L=0, 2, 
\cdots, N/2$ for even $N/2$, and $L=1, 3, \cdots, N/2$ for odd $N/2$).}
\end{table}

\begin{table}[htb]
\begin{center}
\begin{tabular}{c|c|c|c|c|c|c|c|c|c|c|c}
\hline\hline
$N$ &  $L=0$ & 2 & 3 & 4 & 5 & 6 & 7 & 8 & 9 & 10 & 12\\ 
\hline
8 & 0.78 & 0.54 & 0.65 & 0.47 & 0.36  & 0.45  & - & 0.21 & - & - & -\\
10 & 0.67 & 0.48 & 0.49 & 0.47 & 0.21 & 0.34  & 0.26 & 0.32 & - & 0.02  & - \\
12 & 0.42 & 0.32 & 0.27 & 0.32 & 0.17 & 0.28  & 0.21 & 0.23 & 0.23 & 0.24 & 0.07 \\
\hline\hline
\end{tabular}
\end{center}
\caption{\label{PfQHproj4}Overlaps between the PfQH basis for four 
quasiholes and the lowest energy states for Coulomb interaction in the 
second Landau level at $2Q=2N-1$.  The overlap at a given $L$ is defined as
${\cal O}=\sum_{i,j}^{\cal N}|\langle\Psi^{(3)}_{\rm 4-qh, i}|\Psi^{\rm C}_{\rm 4-qh, j}\rangle|^2 / {\cal N}$, where 
${\cal N}$ is the number of degenerate multiplets of $H^{(3)}$ at $L$ \cite{Read1}, 
and $i,j=1,\cdots {\cal N}$.  The states 
$\Psi^{\rm C}_{\rm 4-qh, j}$ represent the ${\cal N}$ lowest energy eigenstates of 
the Coulomb interaction. 
}
\end{table}

Tables \ref{PfQHproj2} and \ref{PfQHproj4} show the 
overlaps between the PfQH basis and the corresponding number of 
lowest energy states for the Coulomb interaction, for two as well as four 
quasiholes.  The overlaps are low by the FQHE standards.  Also, in general, 
no distinct quasihole band analogous to the PfQH band is identifiable in the exact 
Coulomb spectrum.  The Coulomb interaction 
does not simply lift the degeneracy of the PfQH states but changes the structure of 
the low energy sector in a fundamental manner.

\begin{figure}[!htbp]
\begin{center}
\includegraphics[width=\columnwidth, keepaspectratio]{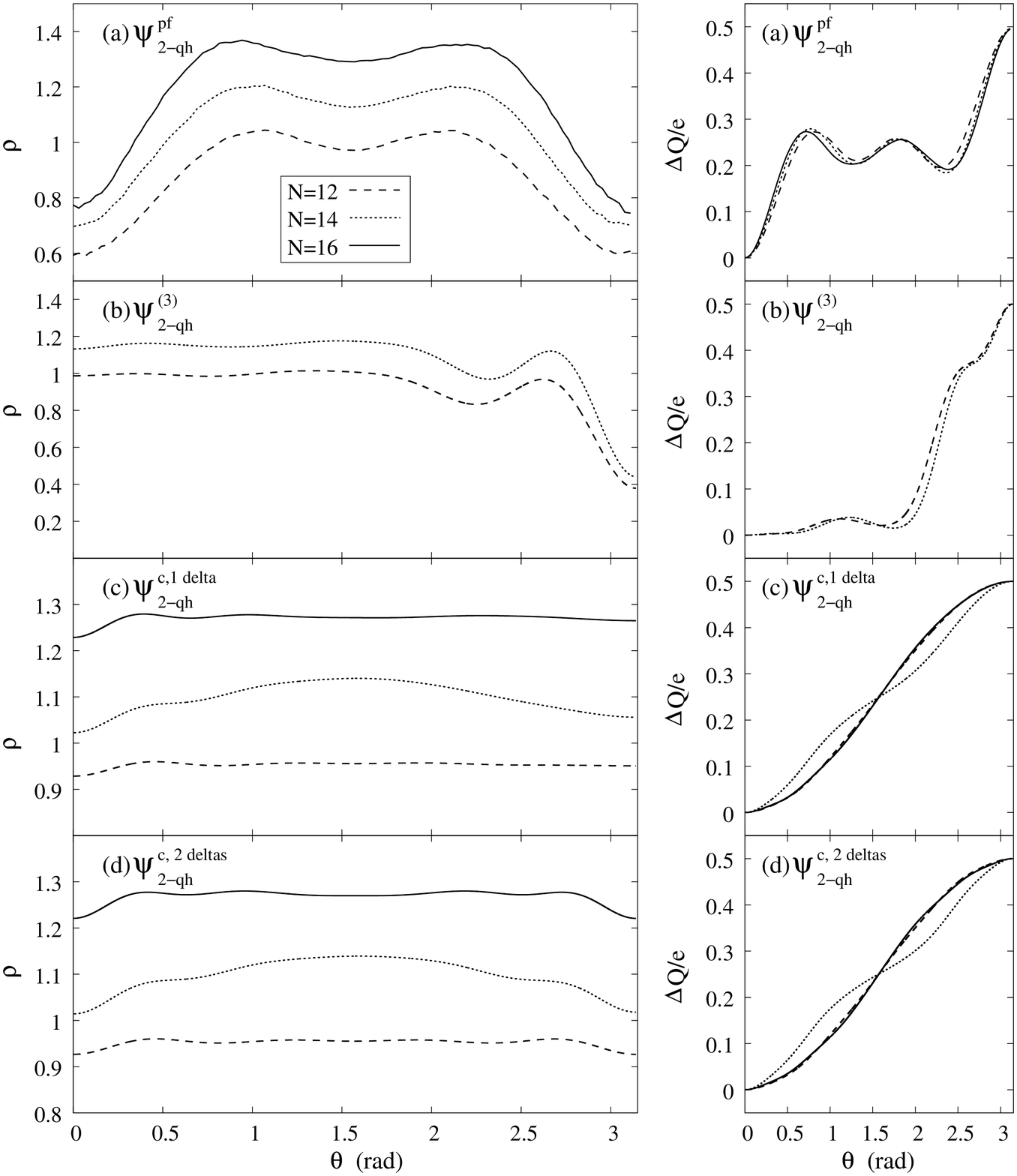}
\end{center}
\caption{Left panel : Charge densities for the ground state for
$\Psi^{\rm Pf}_{\rm 2-qh}$ (a),  $\Psi^{\rm (3)}_{\rm 2-qh}$ (b) and
$\Psi^{\rm C}_{\rm 2-qh}$ (d) in the presence of two delta function
impurities at the two poles, and for $\Psi^{\rm C}_{\rm 2-qh}$ (c)
in the presence of one delta function impurity at the north pole.
The results are shown for $N=12$ (dashed lines), $N=14$ (dotted lines)
and $N=16$ (solid lines) electrons at $2Q=2N-2$.
The density for the Pfaffian wave function $\Psi^{\rm Pf}_{\rm 2-qh}$
(obtained by Monte Carlo) is also shown for comparison,
with the quasiholes placed on the two poles.
When the ground state has $L_z\neq 0$, there are 
two degenerate states at $\pm L_z$; we have shown only one of them for simplicity.
The normalization is chosen to ensure that the integral of the entire sphere gives $N$.
Right panel: the integrated excess charge for each density, normalized so that the 
total charge excess is $1/2$.  Two spatially separated 
quasiholes will exhibit a step at charge 1/4.}\label{density2qh}
\end{figure}

\begin{figure}[!htbp]
\begin{center}
\includegraphics[width=\columnwidth, keepaspectratio]{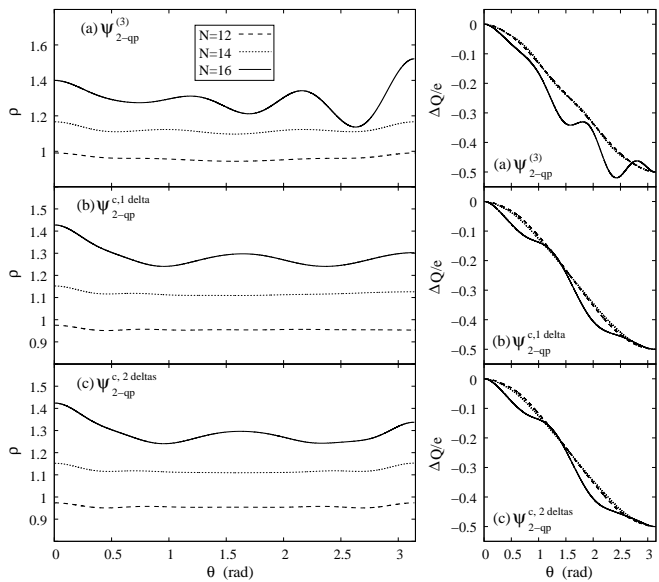}
\end{center}
\caption{Same as in Fig.~\ref{density2qh}, but 
for quasiparticles.  There are only three panels, because a Pfaffian wave function 
analogous to $\Psi_{2-qh}$ is not known.}\label{density2qp}
\end{figure}

We next investigate spatially localized quasiholes and quasiparticles, as needed 
for an evaluation of their braiding phases.
It is natural to attempt to localize them with the help of weak delta function 
potentials of appropriate sign, which can serve 
as a model of an STM tip for manipulating them. 
We have studied a range of strengths for the delta function potential, but we show
results below for weak delta functions of strength $(0.005/\sqrt{Q})e^2/\epsilon l_B$.
We place a delta function at one or both poles of the sphere, 
so $L_z$ continues to be a good quantum number\cite{comment}.

We first consider the $H^{(3)}$ model.  The impurity, being weak, does not cause 
significant mixing with states outside the PfQH sector.   A single delta function impurity 
in the lowest LL localizes a vortex rather than a Pfaffian quasihole for 
the following reason.  The energy 
of a given wave function is equal to a properly weighted average of the densities at 
the positions of the delta functions (for weak impurity strengths).  
For a delta function at $(U,V)$, the lowest 
energy state (which has zero energy independent of the strength of the delta 
impurity) is the one in which {\em both} quasiholes  
localize at $(U,V)$, producing a vortex $\Psi_V$ with vanishing density at $(U,V)$.
(The binding of quasiholes remains valid also for a delta function in the second LL, 
although the density has more complicated structure than a simple vortex, as seen in 
Fig. \ref{density2qh}.)  We ask if it is possible to split the vortex into two 
quasiholes with the help of {\em two} delta function impurities, placed at the two 
poles of the sphere.  The diagonalization is performed in the {\em full} $L_z$ subspace,
including states within and outside the PfQH band; 
for sufficiently weak delta function strengths, however, the solution is 
essentially restricted to the PfQH sector. (For example, for 
$N=14$, the overlap between the ground state with delta functions 
and the corresponding state in the PfQH sector is $0.999$.) The lowest energy eigenstate of 
this model is denoted by $\Psi^{(3)}_{\rm 2-qh}$.   Surprisingly, 
as seen in Fig.~\ref{density2qh}b, two delta impurities fail to separate two quasiholes.
The Pfaffian quasihole wave function $\Psi^{\rm Pf}_{\rm 2-qh}$, in contrast,
has two reasonably well separated quasiholes of charge 1/4 each, as seen in Fig.~\ref{density2qh}a.  
We also show in Fig.~(\ref{density2qp}a) results for two quasiparticles, and 
again see no indication of two separate charge-1/4 objects.

\begin{table}[htb]
\begin{center}
\begin{tabular}{c|c|c}
\hline\hline
$N$ & $|\langle\Psi_{\rm 2-qh}^{\rm Pf}|\Psi_{\rm 2-qh}^{(3)}\rangle|^2$ & 
$|\langle\Psi_{\rm 2-qh}^{\rm Pf}|\Psi_{\rm 2-qh}^{(C)}\rangle|^2$ 
\\ 
\hline
8   & 0 [0.9831(1)*] & 0.326(2)  \\
10  & 0     [0.862(3)*]     & 0  [0.045(3)*]       \\
12 & 0 [0.959(8)*] & 0.19(2) \\
\hline\hline
\end{tabular}
\end{center}
\caption{\label{overpf} Squared overlaps of the Pfaffian quasihole wave function 
$\Psi^{\rm Pf}_{\rm 2-qh}$  (with 
two quasiholes located at the two poles) with the ground states of $H^{(3)}$ and
Coulomb interaction, with a delta impurity at each pole.  
The overlaps are identically zero when the actual ground state has  
$L_z\neq 0$; in those cases, the overlap of $\Psi^{\rm Pf}_{\rm 2-qh}$ with the 
lowest energy
state in the $L_z=0$ sector is also given (in square brackets, marked by an asterisk).
The overlaps are obtained by the Monte Carlo method, with the statistical 
uncertainty given in brackets.}
\end{table}

Table \ref{overpf} shows the overlap of the Pfaffian 
quasihole wave function, $\Psi^{\rm Pf}_{\rm 2-qh}$, with two quasiholes at the 
two poles, with the exact ground eigenstate in the presence of two weak delta function
impurities at the two poles.  When the ground state has $L_z\neq 0$, the 
overlap of $\Psi^{\rm Pf}_{\rm 2-qh}$ with the lowest 
energy state in the $L_z=0$ sector are also shown.

Tables \ref{proj} and \ref{tab:LLz} give insight into how the exact solution of 
$H^{(3)}$, with two impurities, is different from the Pfaffian wave function 
$\Psi^{\rm Pf}_{\rm 2-qh}$, 
Table \ref{proj} shows the projections of $\Psi^{\rm Pf}_{\rm 2-qh}$ with different zero-energy angular momentum eigenstates of $H^{(3)}$. 
The wave function $\Psi^{\rm Pf}_{\rm 2-qh}$ is a linear superposition of PfQH 
states with several low angular momenta. 
The expectation values of $L_z$ and $L$ for the exact two quasihole / quasiparticle
ground states of $H^{(3)}$ are shown in Table \ref{tab:LLz}. The actual ground state of $H^{(3)}$
in the presence of two delta functions, $\Psi^{(3)}_{\rm 2-qh}$, 
has a high angular momentum in the PfQH sector, whereas 
$\Psi^{\rm Pf}_{\rm 2-qh}$ predominantly contains low angular momentum 
components.

\begin{table}[htb]
\begin{center}
\begin{tabular}{c|c|c|c|c|c|c|c}
\hline\hline
$N$ &  $L=0$ & 1 & 2 & 3 & 4 & 5 & 6 \\ 
\hline
8  & 0.33  & -    & 0.65  & -    & 0.02  & -    & - \\
10 & -     & 0.66 & -     & 0.33 & -     & 0.01 & - \\
12 &0.29(3)& -    &0.57(5)& -    &0.13(9)& -    &0.01(1)\\
\hline\hline
\end{tabular}
\end{center}
\caption{\label{proj} Projections of the quasihole wave function 
$\Psi^{\rm Pf}_{\rm 2-qh}$  (with 
the two quasiholes located at the two poles), onto the zero energy angular momentum 
eigenstates of the $L_z=0$ 
PfQH sector, defined as $|\langle \Psi^{\rm Pf}_{\rm 2-qh}|
\Psi^{(3)}_{\rm 2-qh,L}\rangle|^2$.  The angular momentum is $L$, and its 
$z$ component is $L_z=0$.  The projections are evaluated by the Monte Carlo method.
}
\end{table}
  
\begin{table}[htb]
\begin{center}
\begin{tabular}{c|c|c|c|c|c}
\hline\hline
$N$ &  8 & 10 & 12 & 14 & 16  \\ 
\hline
$\Psi^{(3)}_{\rm 2-qh}$   & (4.00,3) & (5.00,4) & (6.00,5) & (7.00,6) & - \\
$\Psi^{\rm C}_{\rm 2-qh}$ & (2.00,0) & (1.01,1) & (0.04,0) & (1.03,0) & (0.08,0) \\
$\Psi^{(3)}_{\rm 2-qp}$   & (0.00,0) & (1.00,0) & (2.00,0) & (3.00,0) & (4.00,2) \\
$\Psi^{\rm C}_{\rm 2-qp}$ & (2.00,0) & (1.00,0) & (0.02,0) & (1.01,0) & (2.01,1)  \\
\hline\hline
\end{tabular}
\end{center}
\caption{\label{tab:LLz} The expectation values of $L$ and $|L_z|$, 
shown as $(L,|L_z|)$,
for the ground states $\Psi^{(3)}_{\rm 2-qh}$, $\Psi^{\rm C}_{\rm 2-qh}$, 
$\Psi^{(3)}_{\rm 2-qp}$,  and $\Psi^{\rm C}_{\rm 2-qp}$ in the presence of two weak 
delta function impurities at the two poles.  
}
\end{table}

We next turn to two Coulomb quasiholes or quasiparticles.  
Here, at first sight, one may expect that even a single delta function should produce 
well separated quasiholes or quasiparticles, 
because it can bind one of them, which then should repel the other.  As seen 
in Figs. (\ref{density2qh}c,d) and (\ref{density2qp}b,c), neither one nor 
two delta functions produce separated 
quasiholes or quasiparticles.  In fact, the charge profile is practically identical for 
the two cases.  The situation is more 
restrictive for the Coulomb interaction because, instead of many degenerate states, 
we have a single ground state multiplet with a definite $L$.  All that {\em weak} 
disorder can do is cause a mixing between the different $L_z$ components 
of the ground state multiplet.  For the case of 
two delta functions at the two poles, $L_z$ is a good quantum number, so the 
delta functions only lift the degeneracy of the $L_z$ states.  
The lack of quasiparticle or quasihole separation in space 
is attributable to the fact that the ground state now has a more or less 
definite $L$.  The absence of exact degeneracy, as found for the $H^{(3)}$ model,
thus inhibits quasihole localization.  The overlaps of the Coulomb quasiholes 
with the Pfaffian quasiholes (Table \ref{overpf}) are very low and rapidly decreasing 
with $N$ \cite{comment2}.

While our study does not rule out well separated charge-1/4 quasiholes for the 
Coulomb problem for larger systems, it is rather striking that no well 
defined quasiholes are 
seen even for systems with as many as 16 electrons, which is sufficiently 
large at least for the {\em Pfaffian} quasiholes to be well separated.   
The Pfaffian model thus fails to  
capture the long range correlations present in the true state.  
Eigenstates with separated quasiholes and quasiparticles do exist
in some of our finite size systems, but 
are not the lowest energy states for our disorder potential.  Separating them 
with more complicated impurity potentials is thus possible, but our 
calculations show that it does not happen generically.

The braiding properties of the Pfaffian quasiholes have been studied numerically by 
Tserkovnyak and Simon\cite{TSimon}.  In view of the above results, it is 
crucial to carry out similar calculations directly for the 
braiding properties of the Coulomb quasiholes and quasiparticles of the 5/2 state.
That, unfortunately, is beyond our present capabilities.
The systems accessible in exact diagonalization study are too small for this 
purpose, as they do not even show charge-1/4 quasiholes.  A computation  
for larger systems would require a knowledge of accurate trial wave functions for 
the Coulomb quasiparticles and quasiholes, not currently available.   
Nonetheless, to the extent that non-Abelian statistics is a consequence of the 
Pfaffian structure, our study questions its validity for the Coulomb problem.
One may ask if the deviation between Pfaffian wave functions and the Coulomb 
solutions is a finite size effect and if the Pfaffian physics can be recovered 
in the thermodynamic limit.  We see no reason for that to happen;  as seen in Tables 
\ref{PfQHproj2}-\ref{overpf}, the Pfaffian wave functions rapidly 
deteriorate with increasing $N$\cite{comment3}.
Larger systems will surely produce many 
{\em quasi} degenerate states for quasiholes, but they are unlikely to 
bear any relation to the zero energy states of the PfQH sector. 
Also, a gap separating ``quasihole states" from rest of the states is crucial 
for maintaining their ``topological"  integrity;  such a gap is present for the $H^{(3)}$ 
model, but not for the Coulomb Hamiltonian.  These considerations suggest that 
the braiding properties of the Pfaffian quasiholes are unlikely to carry 
over to the Coulomb quasiholes.

The existence of charge-1/4 quasiholes is necessary (charge-1/2 vortices have 
abelian braiding statistics) but not sufficient for non-Abelian braiding statistics.  
The braiding statistics is much more sensitive to subtle long range correlations 
than the fractional charge.  As an example, Laughlin's 
wave function for the quasiparticle at 1/3 has the correct charge but  
an incorrect {\em abelian} braiding statistics \cite{Kjonsberg}. 

If the excitations of the 5/2 state are not well described by Pfaffian wave function, 
what describes their physics?
Further work will be needed to answer this question.  The CF theory 
may shed some light on that.  It has been shown \cite{Toke06} that 
the residual interaction between composite fermions opens a gap at $\nu=5/2$.  
In this picture, the quasiparticles are excited composite fermions, although 
heavily renormalized by interaction.

We thank IDRIS-CNRS for a computer time allocation, 
and the High Performance Computing (HPC) group at Penn State University ASET 
(Academic Services and Emerging Technologies)
for assistance and computing time on the Lion-XL and Lion-XO clusters.
JKJ thanks Hans Hansson for stimulating discussions. 
Partial  support of this research by the National Science Foundation under 
grant No.\ DMR-0240458 is gratefully acknowledged.

\newcommand{\PRL}{Phys.\ Rev.\ Lett.}
\newcommand{\PRB}{Phys.\ Rev.\ B}
\newcommand{\NPB}{Nucl.\ Phys.\ B}


\begin{thebibliography}{99}
\bibitem{comment}  The results depend slightly (though not qualitatively) on whether 
the impurity is a delta function in the lowest or the second LL.  With the exception of Fig. 1a, the results in Figs. 1 and 2 depict the electron density in the second LL for a delta function impurity in the second LL.
\bibitem{comment2} Slight modification in the pseudopotentials does not affect 
the picture qualitatively. For two quasiholes at $N=14$, 
changing $V_1$ pseudopotential by $\delta V_1=0.03$, which results 
in the maximum overlap with the PfQH sector, 
produces a ground state which has an overlap of $0.998$ with the pure Coulomb ground 
state.
\end{thebibliography}
\end{document}